# On-site Noise Exposure technique for noise-robust machine fault classification

Wonjun YI[1]; Jung-Woo CHOI[2]

[1,2] School of Electrical Engineering, Korea Advanced Institute of Science and Technology (KAIST),

Republic of Korea

**ABSTRACT**

In-situ classification of faulty sounds is an important issue in machine health monitoring and diagnosis. However, in a noisy environment such as a factory, machine sound is always mixed up with environmental noises, and noise-only periods can exist when a machine is not in operation. Therefore, a deep neural network (DNN)-based fault classifier has to be able to distinguish noise from machine sound and be robust to mixed noises. To deal with these problems, we investigate on-site noise exposure (ONE) that exposes a DNN model to the noises recorded in the same environment where the machine operates. Like the outlier exposure technique, noise exposure trains a DNN classifier to produce a uniform predicted probability distribution against noise-only data. During inference, the DNN classifier trained by ONE outputs the maximum softmax probability as the noise score and determines the noise-only period. We mix machine sound and noises of the ToyADMOS2 dataset to simulate highly noisy data. A ResNet-based classifier trained by ONE is evaluated and compared with those trained by other out-of-distribution detection techniques. The test results show that exposing a model to on-site noises can make a model more robust than using other noises or detection techniques.

## 1. INTRODUCTION

As the manufacturing process is automatized in a smart factory, on-site fault classification based on machine noise is gaining much attention. In a factory, however, noise is always present and the machine produces a sound only during its operation. Consequently, there may be several signal periods where no machine sound is present but only factory noise exists. For robust autonomous fault diagnosis, a fault classifier is required to be robust to a high level of background noise and to distinguish noise-only periods from machine sounds.

Previously, research on out-of-distribution (OOD) detection has been conducted to detect outliers lying outside of the distribution formed by normal data. Such techniques can be utilized to detect noise-only periods in our application. For example, in (1), OOD samples were detected based on the maximum softmax probability predicted from a DNN model trained by a classification task. Another OOD detection algorithm utilizing the free energy function was also studied in (2).

On the other hand, outlier exposure (OE) has been considered an effective way to improve the OOD performance and robustness of the classifier (3, 4). OE utilizes an outlier dataset disjoint from in-distribution and test-time data. In OE, a model is trained to output uniform distributions for outlier data to make maximum softmax probability (MSP) as low as possible. Using the difference in MSPs for the in-distribution and OOD data, the model can detect OOD samples at test time.

In this paper, we report the importance of noise types used for OE. As mentioned, the objective of the training is (a) to classify the fault types and (b) to detect noise-only data. To accomplish these two goals at the same time, we expose a DNN classifier to different types of noises and compare the performance. Like OE, the classifier is trained to produce a uniform probability distribution for noise signals and to predict a one-hot vector for noisy machine sounds. However, we investigate the performance difference when noises used for OE are from similar or dissimilar distribution to those of on-site factory noises embedded in the noisy machine sound signal. We denote the OE with on-site factory noises as the on-site noise exposure (ONE) and compare the performance of the model trained by ONE with models exposed to different types of noises. In addition, to verify the effectiveness of ONE, we also compare it with other training techniques, such as mapping noise-only data into an additional class or free energy-based techniques: energy score and energy-bounded learning (2). The ToyADMOS2 dataset is used to synthesize machine sounds with factory noises and noise-only data,

---

and the noise-only period detection and fault type classification performances are evaluated. The result shows that ONE outperforms other training techniques, as well as the OE using other types of noises.

## 2. BACKGROUND

### 2.1 DNN classifier and softmax score

A DNN classifier $f(\cdot)$ is trained to classify in-distribution data $x_{in} \in D_{in}$ and predicts softmax probability $f(x_{in}) \in \mathbb{R}^K$ for $K$ classes. The training of the classifier is done such that the categorical cross entropy (CCE) between the target probability $y \in \mathbb{R}^K$ given by a one-hot vector and the predicted probability $f(x_{in})$ can be minimized. At inference time, the softmax score $A(x)$ can be calculated by taking the maximum value of the predicted probability

$$A(x) = \text{Max}(f(x)). \tag{1}$$

Here, $\text{Max}(\cdot)$ is the operator taking the maximum value of a vector. In the previous study (1), the softmax score was used as a measure of OOD, i.e., data are regarded as in-distribution if the softmax score is higher than a pre-defined threshold and OOD otherwise.

### 2.2 Outlier exposure

For outlier exposure, a DNN model is trained by not only the in-distribution dataset $D_{in}$ but also an auxiliary dataset $D_{OE}$ for the outliers. The DNN classifier is trained to classify in-distribution data $x_{in}$ as a ground truth one-hot vector $y$ but to produce uniform distribution $u = [1, \cdots, 1]^T / K$ against the outlier data $x_{OE}$. The total loss function $\mathcal{L}_{total}$ can be written as

$$\mathcal{L}_{total} = \mathbb{E}_{(x_{in}, y) \sim D_{in}}(\mathcal{H}(y, f(x_{in}))) + \mathbb{E}_{x_{OE} \sim D_{OE}}(\alpha \mathcal{H}(u, f(x_{OE}))). \tag{2}$$

$\mathcal{H}(\cdot)$ denotes the cross-entropy, so the first and second terms of Eq. (2) represent CCE of in-distribution and outlier data, respectively. The constant $\alpha$ is a balancing weight for the joint training. After training, the softmax score $A(x)$ will be high for in-distribution data $x_{in}$ but low for $x_{OE}$, yielding the estimation of outliers. Although the outlier dataset $D_{OE}$ is different from the actual OOD data used in the test time, the DNN classifier learns the ability to discriminate OOD data by tightening the decision boundary to exclude the outliers provided for OE.

### 2.3 Energy score

Energy score is another measure introduced for OOD detection, which is defined as the negative value of the free energy function:

$$-E(x) = T \log \sum_{k=1}^{K} \exp\left(\frac{g_k(x)}{T}\right), \tag{3}$$

where $g_k(x)$ is the logit value of the classifier for the $k$ th class, and $T$ is the temperature parameter (2). It was known that the softmax score is a special case of the energy score (2), where all logits are biased by the maximum logit value. Since the biased scoring function is not desirable for OOD detection, the energy score is claimed to be more advantageous than the softmax score.

### 2.4 Energy-bounded learning

While the energy score only considers a different scoring of a trained network, the energy-bounded learning (2) utilizes the free energy function for training. The DNN classifier is trained to produce a high energy score for in-distribution data and a low score for outlier data. To this end, the regularization loss $\mathcal{L}_{energy}$ is defined using two squared hinge loss terms as below formula:

$$\mathcal{L}_{energy} = \mathbb{E}_{(x_{in}, y) \sim D_{in}}(\text{Max}(0, E(x_{in}) - m_{in}))^2 + \mathbb{E}_{x_{OE} \sim D_{OE}}(\text{Max}(0, m_{OE} - E(x_{OE})))^2. \tag{4}$$

Here, $m_{in}$ and $m_{OE}$ are margin hyperparameters for in-distribution and outlier data, respectively. The total loss unifying the classification and regularization losses can be described as shown below:

$$\mathcal{L}_{total} = \mathbb{E}_{(x_{in}, y) \sim D_{in}}(\mathcal{H}(y, f(x_{in}))) + \beta \mathcal{L}_{energy} . \tag{5}$$

Here, $\beta$ is a balancing weight for joint training. During inference, an energy score is used for OOD detection.

## 3. ON-SITE NOISE EXPOSURE (ONE)

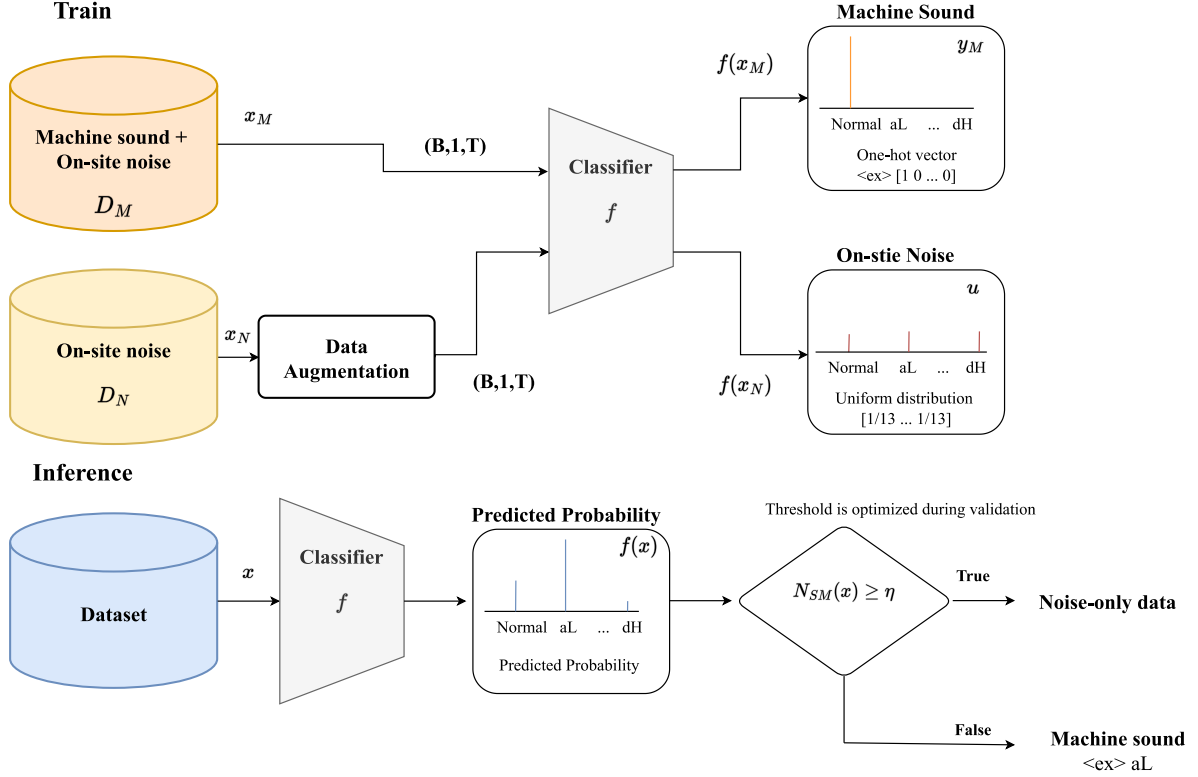

Figure 1 – Training and inference procedures of ONE

In-situ data acquired in a factory can be split into signal segments of a fixed length. Each segment can include only background noises, which we want to determine as outliers. To enhance the detection performance, we apply the OE technique using the background noises of the factory measured while the machine is inactive.

Figure 1 illustrates the training and inference procedures of ONE. The dataset to train the DNN model is the combination of in-distribution data $D_M$ with machine sounds contaminated by on-site factory noises and noise-only data $D_N$ measured during noise-only periods. In the training step, two input batches are sampled from $D_M$ and $D_N$. Both batches have the data size $(B, 1, T)$, where $B$ represents the batch size, 1 is the number of the channel (mono channel), and $T$ represents the number of samples in time. During training, the classifier predicts the fault type probability $f(x_M)$ from the input $x_M \in D_M$, whose target $y_M$ is given by a one-hot vector. Like OE, the classifier is trained to output a uniform distribution for noise-only data $x_N \in D_N$. Similar to Eq. (2), the loss function can be defined as

$$\mathcal{L}_{NE} = \mathbb{E}_{(x_M, y_M) \sim D_M}(\mathcal{H}(y_M, f(x_M))) + \alpha \mathbb{E}_{x_N \sim D_N}(\mathcal{H}(u, f(x_N))) . \tag{6}$$

We denote the second term of Eq. (7) as a *noise exposure loss*.

At the inference time, we determine noise-only data using the *noise score* $N(x)$ defined as

$$N_{SM}(x) = 1 - \text{Max}(f(x)) . \tag{7}$$

The model calculates the noise score for each data at inference time and judges as noise-only data when the noise score is higher than the pre-defined threshold $\eta$. When the noise score is lower than the threshold, the model classifies the fault type by finding the class corresponding to the maximum of $f(x)$. The threshold is determined through the validation step such that the optimal model performance (macro F1 score) can be obtained, and then the fixed threshold is used for the test.

One possible alternative form of noise exposure is to use the free energy function (FE) as the noise score. That is,

$$N_{FE}(x) = -E(x).\tag{8}$$

The energy-bounded learning (EB) mentioned in Eq. (1) can also be utilized for the training with noise exposure. In this case, the loss function can be defined as

$$\mathcal{L}_{EB} = \mathbb{E}_{(x_M, y) \sim D_M}(\mathcal{H}(y_M, f(x_M))) + \beta \mathcal{L}_{energy}\tag{9}$$

where $\mathcal{L}_{energy} = \mathbb{E}_{(x_M, y_M) \sim D_M}(\text{Max}(0, E(x_M) - m_M))^2 + \mathbb{E}_{x_N \sim D_N}(\text{Max}(0, m_N - E(x_N)))^2$,

for margin hyperparameters $m_M$ and $m_N$ of machine sound and noise-only data, respectively. For energy-bounded learning, the same noise score as Eq. (8) is used.

The last alternative form we examine for noise exposure is to train a model to classify the noise-only data into a separate class (additional class; AC). The loss function for this objective can be written as

$$\mathcal{L}_{AC} = \mathbb{E}_{(x_M, y) \sim D_M}(\mathcal{H}(y_M, f(x_M))) + \mathbb{E}_{x_N \sim D_N}(\mathcal{H}(y_N, f(x_N))),\tag{10}$$

where the target probability vector $y_N \in \mathbb{R}^{K+1}$ is equal to one only for the ($K+1$)th class and zeroes otherwise.

## 4. EXPERIMENT

### 4.1 Dataset

Table 1 – The number of data of each class for the toy car dataset

|  | Normal | Fault | Noise |
| --- | --- | --- | --- |
| Train | 200 | 600 (50 per each fault type) | 800 |
| Validation | 100 | 300 (25 per each fault type) | 400 |
| Test | 100 | 300 (25 per each fault type) | 400 |

Table 2 – The number of data of each class for the toy train dataset

|  | Normal | Fault | Noise |
| --- | --- | --- | --- |
| Train | 480 | 1440 (120 per each fault type) | 1920 |
| Validation | 160 | 480 (40 per each fault type) | 640 |
| Test | 160 | 480 (40 per each fault type) | 640 |

For the experiment, we used the ToyADMOS2 dataset (5) consisting of sounds from two toy machines (a toy car and a toy train) and environmental noises. Among five different machine model types and speed levels in the dataset, we selected model type A and speed 1 for both toy car and toy train data. Conditions of each machine are labeled by four types of faults (a, b, c and d) and three damage levels (low, middle, and high), so there is a total of 13 conditions including the normal condition. Every data is 12 s long and the sampling rate is 16 kHz. Noise data of ToyADMOS2 include factory noise signals recorded in four different environments: N1, N2, N3, and N4. We denote these as the *noise environment* to indicate whether the model is trained and tested by the same or different noise data. To increase the diversity of noise data, the time shift (0–2 s) and volume perturbation (0.5–2 times) augmentation was applied to each factory noise signal during the training step.

To simulate noisy environments, we mixed machine sound and noise data with SNR from -10 dB to 0 dB. Considering the rarity of faulty machine data in real situations, we used four times as many normal data as fault data. Details of the train, validation, and test dataset are presented in Tables 1 and 2. In this work, noise-only data are also required for simulating noise-only periods, so we separately prepared the noise dataset by collecting the noise data in ToyADMOS2.

### 4.2 Model architecture

Table 3 – Model structure

| Operation | Input | Output |
| --- | --- | --- |
| Mel spectrogram | $(B, 1, 192000)$ | $(B, 1, 128, 374)$ |
| Batch normalization | $(B, 1, 128, 374)$ | $(B, 1, 128, 374)$ |
| Feature extractor | $(B, 1, 128, 374)$ | $(B, 1024, 1, 1)$ |
| Squeeze | $(B, 1024, 1, 1)$ | $(B, 1024)$ |
| Linear classifier | $(B, 1024)$ | $(B, 13)$ |
| Softmax activation | $(B, 13)$ | $(B, 13)$ |

We modified ResNet for ASC (Acoustic Scene Classification) proposed by Koutini *et al.* (6). The model structure and sizes of input and output are presented in Table 3. The model first transforms a raw audio signal into a Mel spectrogram, which is then standardized by batch normalization and put into the feature extractor. The output of the feature extractor is squeezed and fed into the linear classifier layer. Lastly, softmax activation is applied to generate the predicted probability. Since there are 13 machine condition classes, the output of the model has a shape of $(B, 13)$ for batch size $B$.

### 4.3 Train and validation

We trained the model for 100 epochs. The batch size of machine sound and the noise was 8, respectively, so the total batch size was 16. Adam optimizer was used, and the learning rate was 1e-4 from epoch 1 to epoch 30. Until epoch 90, the learning rate was linearly decreased to 1e-5 and then maintained to epoch 100.

We tested and compared the model trained by ONE with the model only using the softmax score, free energy score (FE), energy-bounded learning (EB), and classifying noise-only periods as an additional class (AC). Details of all techniques are described in Table 4.

Table 4 – Details of techniques

| Techniques | Loss | Noise Score | Noise exposure data |
|---|---|---|---|
| Softmax score | CCE | Negative MSP | Not exist |
| Noise exposure (NE) | CCE, noise exposure loss | Negative MSP | Exist |
| Energy score (FE) | CCE | Negative energy score | Not exist |
| Energy-bounded learning (EB) | CCE, regularization loss | Negative energy score | Exist |
| Additional class with noise data (AC) | CCE | - | Exist |

To construct loss functions, the balancing weight was set to $\alpha = 0.5$ for noise exposure, and $\beta = 0.1$ for energy-bounded learning. The margins of energy-bounded learning, $m_M$ and $m_N$ were set to $-25$ and $-7$, respectively. The temperature parameter $T$ was equal to 1. The model was validated after the end of each epoch. The noise score threshold $\eta$ was set to optimize the model performance and recorded for each epoch. After 100 epochs, the best model parameter and threshold were used for the test.

### 4.4 Test

The evaluation metric used for all experiments is the macro F1 score, which is the average of all F1 scores across 14 different classes (one normal class, one noise class, and 12 fault classes). Since the noise class is included in the classification labels, both the fault-type classification and noise detection performance can be evaluated by a single measure. In detail, the macro F1 score is defined as

$$F1_{Macro} = \frac{1}{14}\sum_{k=1}^{K} F1_k . \qquad (11)$$

Each experiment was repeated three times with different random seeds, and the distribution of scores was presented with its mean value.

## 5. RESULTS

We conducted three different experiments to verify the noise exposure. In section 5.1, we compare ONE with other techniques by training and testing the models using the noises from the same environment. In section 5.2, we examine the performance change when the noises used for the train are different from those for the test. Lastly, in section 5.3, we compare the performance of noise exposure for on-site noise and other-site noise datasets.

### 5.1 Performance under the same noise environment

In the first experiment, we compared five different methods using different on-site noises. The noises from the same environment were used for the train and test datasets but with no overlapping samples between them. Figure 2 shows the macro F1 scores of the comparison sets. We can see that noise exposure (NE) performs best for the toy car. For the toy train, noise exposure outperforms the others for noise environment N1, but energy-bounded learning is better than the others for noise environments N2, N3, and N4. Nevertheless, noise exposure still shows the second-best performance among techniques. A similar trend can be found for the toy train data (Figure 3), where noise exposure and energy-bounded learning show the best performances.

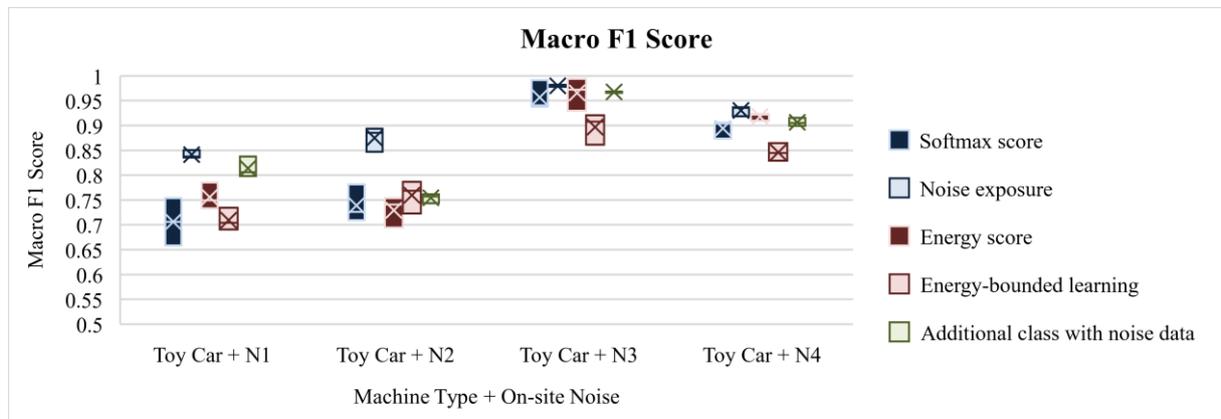

Figure 2 – Performance comparison of different models
when the noises used for the training and test are from the same environment (toy car data).
(N1–N4: noises from different environments)

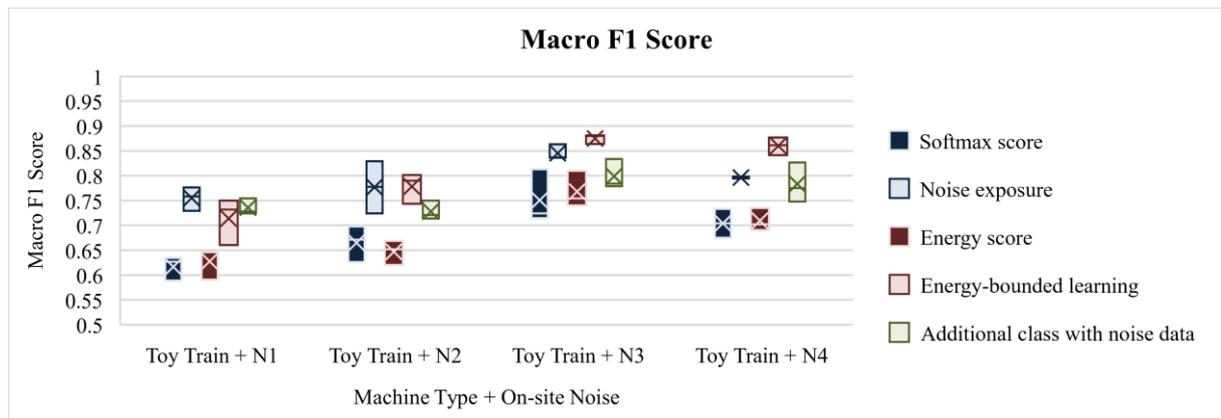

Figure 3 – Performance comparison of different models
when the noises used for the training and test are from the same environment (toy train data).
(N1–N4: noises from different environments)

### 5.2 Performance under the unseen noise environment

Test results when noises from different environments were used for the train and test are presented in Tables 5 and 6. The first cell of each row indicates the noise environment used for train, validation, and test, respectively. Each column represents the applied technique. The bold text highlights the best case for each row. Tables 5 and 6 reveal that noise exposure performs best for most cases of the toy car and toy train. Even when the noise exposure does not score the best, it still shows the second-best performance.

Table 5 – Performance comparison of different models when the noises used for the training (validation) and test are from different environments (toy car data).

| Train Noise / Validation Noise / Test Noise | Softmax score | Noise exposure | Energy score | Energy-bounded learning | Additional class with noise data |
|---|---|---|---|---|---|
| N1 / N1 / N2 | 0.564 | **0.777** | 0.509 | 0.640 | 0.692 |
| N1 / N1 / N3 | 0.490 | **0.676** | 0.463 | 0.594 | 0.617 |
| N1 / N1 / N4 | 0.386 | 0.518 | 0.387 | 0.189 | **0.521** |
| N2 / N2 / N1 | 0.536 | **0.703** | 0.543 | 0.487 | 0.427 |
| N2 / N2 / N3 | 0.657 | **0.799** | 0.644 | 0.690 | 0.668 |
| N2 / N2 / N4 | 0.343 | **0.655** | 0.322 | 0.218 | 0.377 |
| N3 / N3 / N1 | 0.528 | **0.601** | 0.575 | 0.293 | 0.489 |
| N3 / N3 / N2 | 0.534 | **0.610** | 0.553 | 0.336 | 0.611 |
| N3 / N3 / N4 | 0.380 | **0.397** | 0.392 | 0.240 | 0.324 |
| N4 / N4 / N1 | 0.483 | **0.539** | 0.512 | 0.387 | 0.348 |
| N4 / N4 / N2 | 0.576 | **0.590** | 0.587 | 0.361 | 0.476 |
| N4 / N4 / N3 | **0.579** | 0.577 | 0.551 | 0.377 | 0.522 |

Table 6 – Performance comparison of different models when the noises used for the training (validation) and test are from different environments (toy train data).

| Train Noise / Validation Noise / Test Noise | Softmax score | Noise exposure | Energy score | Energy-bounded learning | Additional class with noise data |
|---|---|---|---|---|---|
| N1 / N1 / N2 | 0.572 | **0.779** | 0.534 | 0.628 | 0.683 |
| N1 / N1 / N3 | 0.589 | **0.769** | 0.591 | 0.587 | 0.687 |
| N1 / N1 / N4 | 0.475 | **0.734** | 0.435 | 0.586 | 0.609 |
| N2 / N2 / N1 | 0.493 | **0.583** | 0.483 | 0.415 | 0.443 |
| N2 / N2 / N3 | 0.575 | **0.694** | 0.566 | 0.551 | 0.574 |
| N2 / N2 / N4 | 0.441 | **0.683** | 0.442 | 0.541 | 0.551 |
| N3 / N3 / N1 | 0.445 | 0.497 | 0.412 | **0.521** | 0.368 |
| N3 / N3 / N2 | 0.418 | **0.520** | 0.407 | 0.519 | 0.414 |
| N3 / N3 / N4 | 0.528 | **0.720** | 0.523 | 0.541 | 0.557 |
| N4 / N4 / N1 | 0.367 | **0.433** | 0.379 | 0.311 | 0.262 |
| N4 / N4 / N2 | 0.429 | **0.537** | 0.424 | 0.470 | 0.393 |
| N4 / N4 / N3 | 0.420 | **0.519** | 0.429 | 0.367 | 0.389 |

### 5.3 Performance of noise exposure using on-site noise and other-site noise

In this experiment, we used different noise datasets for the train, validation, and test. The first columns of Tables 7 and 8 indicate the noise used in the train and validation step, while the other columns represent the macro F1 scores when tested by machine sounds mixed with noises indicated on the first cell of the corresponding column. The noise environment of noise-only data for the test was the same as that of machine sound data for the test. The diagonals of Tables 7 and 8 (grey color) hence indicate the results of on-site noise exposure, and off-diagonals show the results of other-site noise exposure. The test with the toy car and toy train dataset demonstrates that the on-site noise is more beneficial than the other-site noises for the noise exposure. One exceptional case (test with N3) exists in the toy car test, but in general, the model exposed to similar types of noises is more robust to the noisy data. This may seem to be an obvious conclusion but also stresses that we can robustly train a model by exposing it to in-situ noises measured in the factory without the need of collecting various noise data.

Table 7 – Performance comparison of different models trained by noise exposure with on-site (ONE) and other-site noises (toy car data).

| Noise data used for noise exposure | Machine sound with N1 | Machine sound with N2 | Machine sound with N3 | Machine sound with N4 |
|---|---|---|---|---|
| N1 | **0.841** | 0.865 | 0.985 | 0.925 |
| N2 | 0.773 | **0.875** | 0.988 | 0.919 |
| N3 | 0.701 | 0.778 | 0.980 | 0.889 |
| N4 | 0.704 | 0.792 | 0.982 | **0.930** |

Table 8 – Performance comparison of different models trained by noise exposure with on-site (ONE) and other-site noises (toy train data).

| Noise data used for noise exposure | Machine sound under N1 | Machine sound under N2 | Machine sound under N3 | Machine sound under N4 |
|---|---|---|---|---|
| N1 | **0.755** | 0.761 | 0.801 | 0.752 |
| N2 | 0.732 | **0.777** | 0.776 | 0.778 |
| N3 | 0.668 | 0.705 | **0.845** | 0.743 |
| N4 | 0.625 | 0.714 | 0.743 | **0.796** |

## 6. CONCLUSIONS

In this paper, we exposed a DNN model to various noises to build a noise-robust classifier and detect noise-only data. We compared five different outlier exposure methods for this objective. With noise exposure, the classifier was trained to classify machine conditions while producing a uniform predicted probability for noise-only data. The comparison of outlier exposure methods shows that noise exposure has the best or second-best performances irrespective of the types of machine sounds and noises. We also tested noise exposure using different on-site and other-site noises. For the on-site noise test, the same noise dataset split for train and test was employed, whereas different datasets were used for the other-site noise test. Results show that noise exposure with on-site noise outperforms the other-site noise test. These results demonstrate that the noises measured in the same environment where the machine operates can be used as a noise exposure dataset and can improve the robustness of fault classifiers.

## ACKNOWLEDGEMENTS

This work was supported by the National Research Foundation of Korea (NRF) grant funded by the Korean Government (Ministry of Science and ICT) (No. NRF-2020M2C9A1062710) and supported by the BK21 Four program through the National Research Foundation (NRF) funded by the Ministry of Education of Korea.

## REFERENCES

1. Hendrycks D, Gimpel K. A Baseline for Detecting Misclassified and Out-of-Distribution Examples in Neural Networks. In: 5th International Conference on Learning Representations, ICLR 2017, Toulon, France, April 24-26, 2017, Conference Track Proceedings.
2. Liu W, Wang X, Owens J, Li Y. Energy-based Out-of-distribution Detection. In: Advances in Neural Information Processing Systems. Curran Associates, Inc.; 2020. p 21464–75.
3. Hendrycks D, Mazeika M, Dietterich TG. Deep Anomaly Detection with Outlier Exposure. In: 7th International Conference on Learning Representations, ICLR 2019, New Orleans, LA, USA, May 6-9, 2019.
4. Primus P, Zwifl M, Widmer G. CP-JKU Submission to DCASE'21: Improving Out-of-Distribution Detectors for Machine Condition Monitoring with Proxy Outliers & Domain Adaptation via Semantic Alignment. DCASE2021 Chall. 2021;
5. Harada N, Niizumi D, Takeuchi D, Ohishi Y, Yasuda M, Saito S. ToyADMOS2: Another dataset of miniature-machine operating sounds for anomalous sound detection under domain shift conditions.
6. Koutini K, Eghbal-zadeh H, Dorfer M, Widmer G. The Receptive Field as a Regularizer in Deep Convolutional Neural Networks for Acoustic Scene Classification. In: 2019 27th European Signal Processing Conference (EUSIPCO). A Coruna, Spain: IEEE; 2019. p 1–5.